\documentclass[10pt,conference]{IEEEtran}

\usepackage{amsmath,amssymb,amsfonts}
\usepackage{algorithmic}
\usepackage{graphicx}
\usepackage{xspace}
\usepackage{focus} 
\usepackage{stmaryrd}

\newcommand{\focust}{\textsc{Focus}$^{ST}$}
\newcommand{\focuse}{\textsc{Focus}$^{E}$}

\begin{document}

\title{\focuse: A semantic extension of \focust}
\author{\IEEEauthorblockN{
        Maria Spichkova
    }
    \IEEEauthorblockA{
        RMIT University}
}
\maketitle

\begin{abstract}
To analyse and verify the safety and security properties of interactive systems, a formal specification might be necessary. There are many types of formal languages and frameworks. The decision regarding what type of formal specification should be applied in each particular case depends on many factors. One of the approaches to specify interactive systems formally is to present them as a composition of components processing data and control streams. 
In this short paper, we present \focuse, a formal approach for modelling event-based streams. The proposed approach is based on a formal language \focust\, and can be seen as its semantic extension.  
\end{abstract}

\maketitle

\section{{Introduction}}

A formal specification might be necessary to analyse and verify the safety and security properties of interactive systems, if we would like to achieve a higher level of quality assurance than might be achievable using testing-only approaches. While testing allows to check whether the system behaves correctly for a given point-wise scenario, verification allows us to check properties of the system. 
There are many types of formal languages and frameworks. The choice of what exactly formal language/frameworks should be applied in each particular case depends on many factors: what kind of system we are aiming to specify, what kind of properties we intend to analyse, etc. 

In the case of interactive systems, one of the established approaches for formal specification and verification is to present the system as a composition of components, processing data and control streams, i.e., to apply the principle of modularity. However, the question is how exactly the streams should be specified, as the choice of the specification approach would impact the readability and verifiability of the specification.

Czepa et al.~\cite{czepa2018understandability} conducted experiments with 216 participants to study the understandability of Linear Temporal Logic (LTL,\cite{pnueli1977temporal}),  
Property Specification Patterns (PSP,\cite{dwyer1999patterns}),  
and Event Processing Language (EPL,\cite{wu2006high,paschke2010homogeneou}). Their study demonstrated that  PSP, which is a highly abstract specification language, is generally easier to understand than LTL and EPL.

In this work, we would like to propose an 
a formal approach for modelling event-based streams, which we call \focuse. The proposed approach is based on a formal language \focust\, and can be seen as its semantic extension.   
\focust allows the creation of concise but easily understandable specifications and is appropriate for the application of the specification and proof methodology presented in our previous works.

\section{Background: $Focus^{ST}$}

The \focust\cite{spichkova2014modeling} language was inspired by \Focus\cite{focus,spichkova2017auto},
a framework for formal specification and development of interactive systems.
In both languages, specifications are based on the notion of \emph{streams}.
However, in the original \Focus  input and output streams of a component are mappings
 of natural numbers $\Nat$ to single messages,
whereas a \focust\  stream %
 is a mapping from $\Nat$ to lists of messages within the corresponding time intervals.
 Moreover, the syntax of \focust\ is particularly devoted to specifying spatial (S) and timing (T) aspects in a comprehensible fashion,
which is the reason to extend the name of the language by $^{ST}$.
The \focust\ specification layout also differs from the original one: it
is based on human factor analysis within formal methods~\cite{spichkova2014we,hffm_spichkova,Spichkova2013HFFM}.

\focust\ allows to create concise but easily understandable specifications
and is appropriate for application of the specification and proof methodology presented in~\cite{spichkova}.
This methodology allows to specify components in a way that carrying out proofs is quite simple and scalable to practical problems.
In particular, a specification of a system can be translated to a Higher-Order Logic and verified by the interactive semi-automatic theorem prover Isabelle~\cite{npw}
also applying its component Sledgehammer~\cite{Sledgehammer} that 
employs resolution-based~\cite{Sledgehammer,SledgehammerSpass}
first-order automatic theorem provers (ATPs)
and satisfiability modulo theories (SMT)
solvers to discharge goals arising in interactive proofs. 

\focust\ has been successfully applied in a number of case studies, e.g.,  for analysis of cryptographic properties~\cite{spichkova2018focusst}, analysis of FlexRay~\cite{spichkova2017formal} and CAN~\cite{spichkova2019towards} protocols.

In \focust\, we specify every component using assumption-guarantee-structured templates.
This allows us to avoid the omission of unnecessary assumptions about the system's environment since a specified component is required to fulfil the guarantee only if its environment behaves in accordance with the assumption.
In a component model, one often has transitions with local variables that are not changed.
Also, outputs are often not produced,
e.g., when a component gets no input or some preconditions necessary to produce a nonempty output are violated.
In many formal languages, this kind of invariability has to be defined explicitly to avoid
under-specified component specifications.
To make our formal language better understandable for programmers, we use in
\focust\ so-called \emph{implicit else-case} constructs.
That means, if a variable is not listed in the guarantee part of a transition, it implicitly keeps its current value.
An output stream not mentioned in a transition will be empty.
Further, we do not require introducing auxiliary variables explicitly:
The data type of a not introduced variable is universally quantified in the specification such that it can be used with any data value. 
The \focust\ specifications are a special form of timed automata, so-called \emph{Timed State Transition Diagrams} (TSTDs), which 
 can be described in both diagram and textual form.

 \begin{figure*}[ht!]
\begin{center}
\includegraphics[scale=0.75]{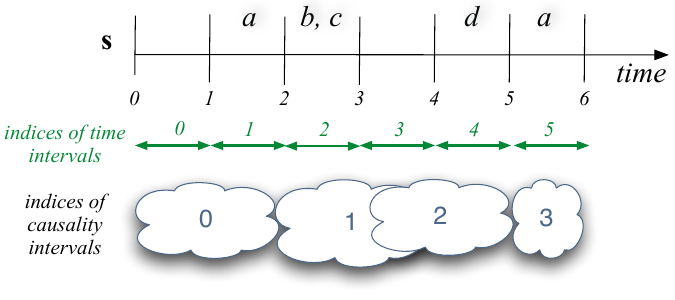}
\end{center}
\caption{Time intervals vs. causality intervals}
\label{fig:ti_ci}
\end{figure*}

 \section{Event-based streams in \focuse}
 
To deal with event-based systems, we suggest inheriting the  \focust syntax accompanied with different semantics. 
For simplicity, we call the new version of the language \focuse.
In  \focuse,  input and output streams of a component are mappings
 of natural numbers $\Nat$ to lists of messages, like in \focust. 
 However, in \focuse\ these lists represent not the messages within the corresponding \emph{time intervals}, but messages within the same 
 \emph{causality intervals}. 
 We can see causality intervals as an abstract view of time intervals (cf. also Figure~\ref{fig:ti_ci}): 
 \begin{itemize}
 \item
If messages belong to the same causality interval,  this means that, according to the system's clock, these messages come simultaneously. 
  \item
If some message $a$ belongs to the causality interval $i$ (from the timed point of view, it belongs to some time interval $t$), 
 and some message $b$ belongs to the causality interval $i+1$, this does not necessary mean that $B$ should belong to the time interval $t+1$, 
 because the causality property insure only the fact ``event $b$ happens after event $a$''.
 Thus, $b$ should belong to the time interval $t+\delta$, where $\delta > 0$. 
 \end{itemize}
 In a special case, the causality intervals of a stream can be equal to the time interval of this stream.
 
 Each message can be seen as a \emph{single event}, but we can also have an additional view on the streams, 
 where a set of messages (single events) from the same causality interval can be denoted as a \emph{combined event}. 
Let us name some of the operators used on \focust 
to specify time intervals in our streams:
$\nempty$ denotes an empty list, e.g., a single time interval without any events,
and $\angles{x}$ a list consisting of the element $x$; %
$\nft{l}$ describes the first element of a list $l$; %
$\ti{s}{t}$	represents the $t$th time interval of the stream $s$. 
We suggest the following notation for \focuse an additional operator 
$\ci{s}{i}$	to represent   the $i$th time interval of the stream $s$. 
For the example presented on Fig.\ref{fig:ti_ci}, we have that\\
\indent $\ti{s}{0} = \nempty$, $\ti{s}{1} = \angles{a}$, $\ti{s}{2} = \angles{b,c}$, $\ti{s}{3} = \nempty$, $\ti{s}{4} = \angles{d}$, $\ti{s}{5} = \angles{a}$, and \\
\indent $\ci{s}{0} = \angles{a}$, $\ci{s}{1} = \angles{b,c}$, $\ci{s}{2} = \angles{d}$, $\ci{s}{3} = \angles{a}$.
\\
\noindent
From purely syntactical point of view, if we take a timed (\focust) stream and remove all empty timed intervals from it, we obtain an event (\focuse) stream. 
Correspondingly, we can define operators over events' causality, e.g., to denote that the $i$th causality interval of a stream $s_{1}$ 
occurs before the $j$th causality interval of a stream $s_{2}$, to denote that events of some type should occur in the stream always 
before some instances of messages of another type, etc.


 \bibliographystyle{plain}

\end{document}